\documentclass[floatfix,secnumarabic,amssymb,nobibnotes,nofootinbib,aps,pra,showpacs]{revtex4-1}
\usepackage{epsfig}
\usepackage{amssymb}
\usepackage[colorlinks=true, citecolor=red, linkcolor=blue, urlcolor=blue]{hyperref}
\usepackage{times}
\usepackage{amssymb}
\usepackage{amsmath}
\usepackage{cleveref}
\usepackage{mathrsfs}
\usepackage{graphicx}
\usepackage{epsfig}
\usepackage{dcolumn}
\usepackage{color}
\usepackage{bm}
\usepackage{graphics,psfrag}
\usepackage{graphicx,psfrag}
\usepackage{braket}
\usepackage{soul}
\usepackage{subfigure}
\newcommand{\be}{\begin{equation}}
\newcommand{\ee}{\end{equation}}
\newcommand{\bea}{\begin{eqnarray}}
\newcommand{\eea}{\end{eqnarray}}
\newcommand{\ba}[1]{\begin{array}{#1}}
\newcommand{\ea}{\end{array}}

\usepackage{graphicx,amsmath,amssymb,bbm,xcolor,hyperref}

\begin{document}
 \title{Trapped ion-mediated interactions between two distant trapped atoms}
 \author{Subhra Mudli$^{\dagger}$, Subhanka Mal$^{\dagger}$, Anushree Dey and Bimalendu Deb}
 \email{msbd@iacs.res.in}
 \thanks{\\ \hspace{-0.5in}$^\dagger$These authors contributed equally to this work.}
 \affiliation{School of Physical Sciences, Indian Association for the Cultivation of Science, Jadavpur, Kolkata 700032, India.}
\begin{abstract}
We theoretically show that when two largely separated trapped atoms interact with a trapped ion via Rydberg excitation of the atoms, the ion-mediated interaction between the atoms exceeds the direct atom-atom interaction by several orders of magnitude. Since the motion of the atoms is much slower than the motion of the ion, we resort to Born-Oppenheimer approximation to calculate the ion-mediated adiabatic potential. We also calculate the ion-mediated phonon modes of the atoms that are separated by more than 10 micron.   For cylindrical geometry of the system and both the atoms being excited to the same Rydberg state, the stretched and center-of-mass (COM) axial or transverse phonon modes are found to be almost degenerate, while the phonon modes are non-degenerate when one atom is in a Rydberg state and the other in the ground state. We discuss the non-adiabatic effects in the system that give rise to a Gauge structure and associated geometric phase in the system. This study may open a new perspective in quantum computing and exploring molecular physics associated with a conical intersection using an ion-atom hybrid architecture.

\end{abstract}

\maketitle

\section{Introduction}

Manipulating coupling or interactions between ionic or atomic qubits which are separated on the micrometer scale is an important step towards quantum information processing in  an ion- or atom-based quantum computer. A two-qubit quantum gate operation \cite{Cirac:PRL:1995,Sorenson:PRL:1999} in ion  traps makes use of the ion-ion Coulomb interaction that, in combination with the harmonic  motion of the trapped ions,  leads to the phononic coupling between the ions. This is accomplished by addressing two ionic qubits individually with Raman laser pulses that control the phononic coupling in a deterministic way.  Since the beam waist of a laser is typically of the micrometer scale, the separations between the qubits in a quantum computing architecture must be larger than a micrometer in order to be able to address the qubits individually with lasers. Two neutral atoms in electronic ground or low lying excited states  generally interact  with a range of  sub-nanometer  scale. This rules out the possibility of any two-qubit gate operation in such neutral atoms using Raman pulses. However, when the atoms are excited to high Rydberg states with the principal quantum number $n \sim 100$, they can interact strongly even at separations on the micrometer scale.  Thus Rydberg excitation provides a promising route to neutral atom-based two- or multi-qubit quantum gate operation. 

In recent times,  Rydberg atoms in optical lattices or optical tweezers have emerged  as viable architecture for neutral atom-based quantum computation and quantum simulation \cite{Saffman:RMP:2010,Beterov:JPB:2016,Browaeys:NatPhys:2020}. Based on Rydberg Blockade  which forbids excitation of a second atom when the first atom is already excited in a Rydberg state provided two atoms are separated not more than a critical distance known as Rydberg radius, Graham {\it et al.} have experimentally demonstrated a two-qubit quantum gate using neutral atoms in an optical lattice \cite{Graham:PRL:2019}. Several quantum algorithms have been implemented on a two-dimensional array of hyper-fine-qubit neutral atoms using Rydberg excitation and individual addressing of single atoms \cite{Graham:Nat:2022}. 

With these current advancements, the arrays of neutral atom qubits in tailor-made optical lattices or optical tweezers have acquired almost all the essential features to serve as a viable quantum computer or a quantum simulator. Over the last two decades, a lot of improvements have been made in design and functionality of the
linear and planner ion traps for applications in quantum computing  and simulation \cite{Leibfried:RMP:2003,Monroe:RMP:2021}. With current pace of progress in both neutral atom- and ion-trap technologies, it is expected that in near future a hybrid quantum architecture combining both trapped ions and trapped atoms will be developed for all or certain tasks in quantum computation and quantum simulation. In fact, a lot of experimental and theoretical research are currently being pursued to understand ion-atom collision physics either in a hybrid trap \cite{Tomza:RMP:2019,Niranjan:Atoms:2021,Eberle:JPCS:2015,Jyothi:RSI:2019,Bahrami:arxiv:2023,Cui:arxiv:2023} or merging an  atom-trap containing ultracold atoms with an ion-trap containing one or a few ions \cite{Ewald:PRL:2019}. The purpose of these studies is to understand the basic elastic and inelastic processes in ion-atom collisions \cite{Rakshit:PRA:2011,Tomza:RMP:2019} at different energy scales and eventually to go to the quantum regime of ion-atom collisions where a few low lying partial waves become important.

Here we address a different question regarding an ion-atom hybrid quantum platform. We consider a scheme where  single atoms in optical tweezers can be brought  in the vicinity of a trapped ion with relatively large separations to avoid any direct atom-ion collision, yet there should be significant ion-atom interaction through Rydberg excitation of the atoms \cite{Secker:PRA:2016,Secker:PRL:2017,Ewald:PRL:2019}. The trapping frequencies of an ion-trap differs from that of an atom-trap typically by two to three orders of magnitude. Consequently, the length scales of these two types of traps differ by one or two orders of magnitude. Ion traps are extraordinarily stable (hours or days) while atom traps have limited life time. The coherence time of an ion qubit is of the order of a second while that of an atomic qubit is typically of 100 microsecond or millisecond order. The first question that comes to mind is then how one can symbiotically combine these widely different energy, length and time scales so that one can form a stable hybrid structure. In fact, the difficulties and possible routes to integrating a single Rydberg atom in an optical trap with a single ion in a Paul trap for  ion-atom two-qubit quantum gate operation  have been theoretically addressed by Secker {\it et al.} \cite{Secker:PRA:2016}.  Setting aside the engineering aspect of constructing a stable hybrid platform combining both ion  and atom traps, here we focus on certain basic physics part of the problem theoretically.  Our purpose is to understand ion-mediated atom-atom
interactions and phononic couplings between the atoms. The long term goal of this study is to use these phononic couplings for fast two qubit quantum gate operations using atomic qubits while ionic qubits may be used as quantum memory.

We consider a model system where two neutral atomic qubits   confined in two similar but largely separate optical tweezers  simultaneously interact  with an ion in a trap.  The center of the ion-trap lies  between the two tweezers' centers as schematically shown in Fig.\ref{fig 1}. One may assume that both the optical tweezers operate at a magic wavelength for which one electronic ground-state and one Rydberg state of the atom can be simultaneously confined in the tweezers. Otherwise, one of the ground-state hyper fine qubit states of the atom may be admixed with a Rydberg state so that the atoms interact with the ion and between themselves primarily through Rydberg excitation.  Since the harmonic trapping frequency of an ion-trap is larger than that of  optical tweezers typically  by  three orders of magnitude, the motion of the ion is much faster than that of the atoms.  Because of the Rydberg excitation-induced long-range atom-ion \cite{Ewald:PRL:2019} and atom-atom interactions \cite{Beguin:PRL:2013} and much faster center-of-mass (COM) motion of the ion than that of the atoms, an intriguing interplay between quantum dynamics of atoms and the ion occurs. 

Here we study ion-mediated interactions and  quantum correlations between the atoms. We use Born-Oppenheimer (BO) approximation \cite{Born:AdP:1927} well-known in molecular physics to clearly separate out the fast ionic degrees-of-freedom (DOF) from the much slower atomic DOF to derive ion-mediated interatomic BO potentials. We consider that the atomic ground state $|g_a \rangle$  is coupled to a low-lying  Rydberg state $|r \rangle = |n, S \rangle $ (where $n$ is the principal quantum number and $S $ refers to the electronic orbital momentum $\ell = 0$) by a two-photon process via an intermediate $P (\ell = 1)$ state. We have chosen $S$ Rydberg state because we want to get rid of  the first order Stark shift on the Rydberg state due to the electric field produced by the ion. At long separations,  the ion-atom interaction potential is of the form $- C_4/r^4$ where $C_4$ is a coefficient proportional to polarizability of the atom due to the ionic field.  There will be a small Stark shift of the $S$ level due to the electric field of the Paul trap electrodes, but we assume that the tweezers'  centers are placed at positions where electric field due to the ion-trap electrodes is almost homogeneous over the extent of the atom's motion in the tweezers. This is necessary to nullify or minimize  any effect of electric field-induced forces on  the atoms.  The Stark-shifted $S$ Rydberg state remains well isolated from all other Stark-shifted atomic levels at a separation much larger than 0.5 $\mu$m \cite{Secker:PRA:2016}.  When the atom is in $|g_a \rangle$ there is practically no interaction between the atom and the ion if the separations between them  are large ($> 0.5 \mu$m).  Since the induced atomic polarizability scales as $n^7$ \cite{Kamenski:JPB:2014}, even with low-lying or moderate Rydberg excitation with  $20 < n < 40$,  the range of the  ion-mediated potentials between the two trapped Rydberg atoms is  greater than the direct Rydberg-Rydberg interaction and the range can be as large as tens of microns. In the leading order, the ion-mediated Rydberg-Rydberg interaction goes as $- 1/r^4$,  and for separations typically larger than 0.5 $\mu $m, it   exceeds the direct interaction between two $S$-state Rydberg atoms  by several orders of magnitude. Remarkably, the ion-mediated long-range interaction between the two atoms depends on atomic internal state. For simplicity, we consider  the ion and the two atoms are harmonically trapped in their respective traps.  We  first diagonalize the  Hamiltonian of atom-ion-atom system considering the atomic COM motion is frozen under BO approximation.  We then calculate axial and transverse phonon modes of the atoms due to ion-induced small oscillations of the equilibrium positions of the two atoms. The non-adiabatic coupling in the system leads  atom-ion-atom entangled states.  The two atoms may be separated by a distance as large as 10 $\mu$m, still they can be entangled due to atoms' internal state-dependent phonon modes. Our study may open a new perspective for quantum computation and simulation using  an atom-ion-atom hybrid quantum node.

The paper is organized in the following way. In Sec.\ref{Sec:2} we present  our model, formulate the problem and obtain analytical solutions. 
In Sec.\ref{Sec:3} we illustrate the numerical results for a typical system considering experimentally feasible system parameters. Finally we conclude in Sec.\ref{Sec:4}. 

\section{ The model and formulation of the problem}\label{Sec:2}
\begin{figure}
  \begin{center}
    \includegraphics[height=2.6in,width=4.5in]{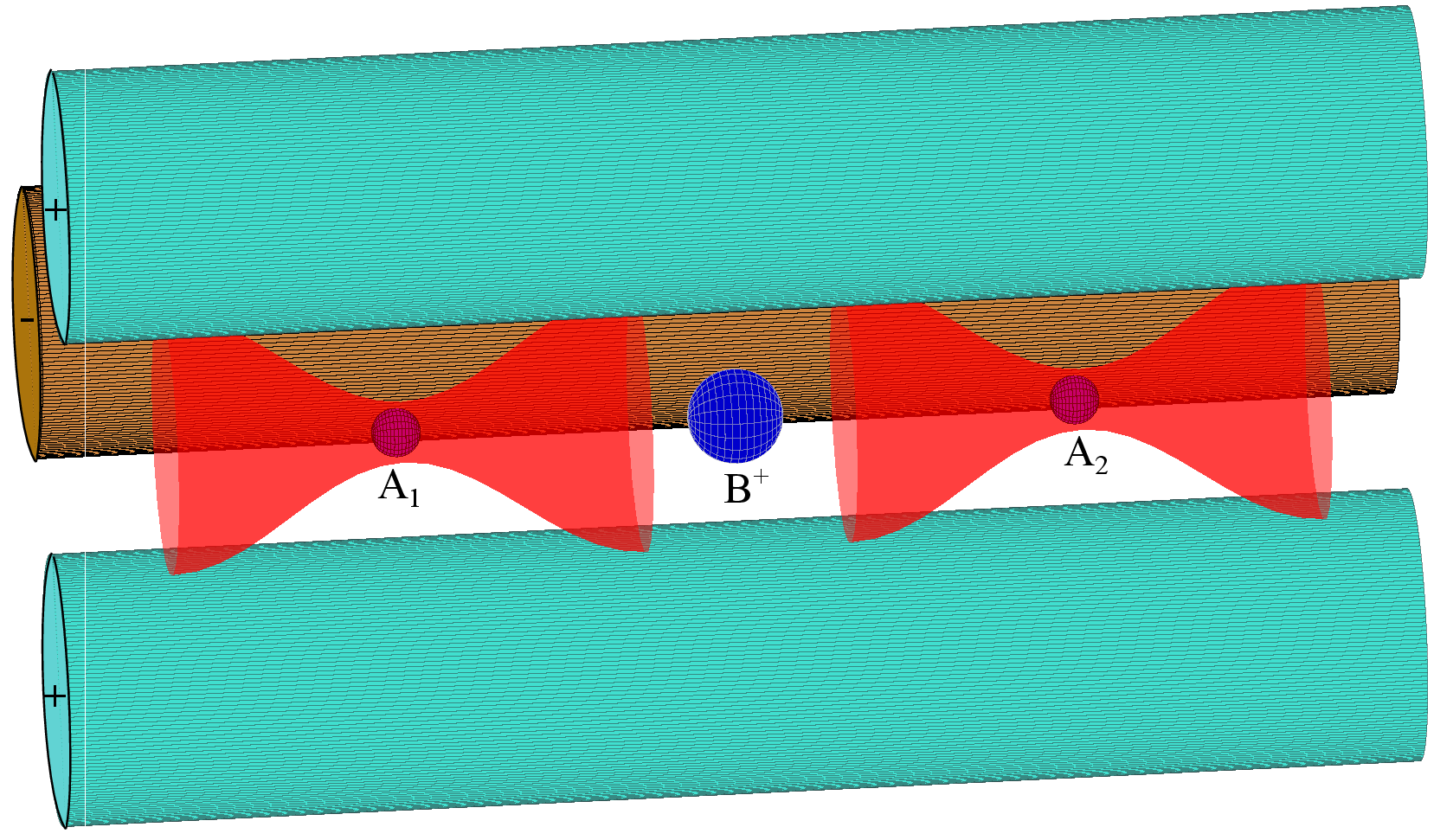}\\
    \includegraphics[height=1.6in,width=5.2in]{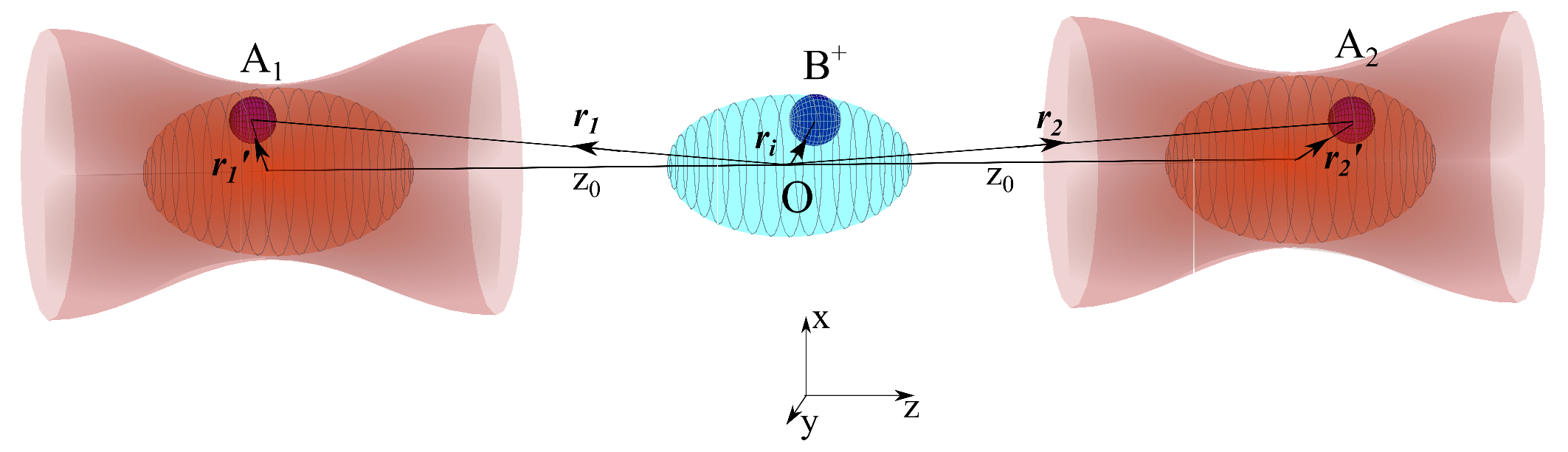}
  \end{center}
  \caption{A schematic diagram of the model system: two  atoms $A_{1}$ and $A_{2}$ trapped in two largely separate optical tweezers simultaneously interact with a single ion $B^{+}$ trapped at the center of a linear Paul trap. Elimination of much faster ionic degrees-of-freedom under Born-Oppenheimer approximation  leads to an effective ion-mediated long-range atom-atom interaction and geometric or Berry phase in the system (see text).}
  \label{fig 1}
\end{figure}
A schematic diagram  of our proposed model is shown in Fig.\ref{fig 1}.  Two identical optical tweezers containing identical single atoms $A_{1}$ and $A_{2}$ of mass $m_a$ are placed symmetrically on the two sides of a single ion $B$ of mass $m_i$ trapped in a Paul trap. The origin of the coordinate system is taken at the electric potential minimum of the ion-trap or the equilibrium position of the ion. Note that other geometries are  also possible, for instance one can consider a planar ion trap or an ion chip  and the optical tweezers may be placed above the surface of the trap.   
To begin with, let us assume that the atoms $A_{1}$ and $A_{2}$ are in the internal (electronic) state $|\alpha \rangle$ and $|\beta \rangle$ respectively; where $\alpha$ and $\beta$ represent ground $\left(g\right)$ or Rydberg state $\left(r\right)$. Since the electric field of the Paul trap electrodes primarily influences the internal electronic states of the atoms, we here do not consider such electric-field-induced effects. The  Hamiltonian of the system is
\begin{eqnarray}\label{eq:1}
    \hat{H}^{\alpha\beta} = \hat{H}^{\alpha\beta}_{{\rm atom}} + \hat{H}^{\alpha\beta}_{\rm{ion-atom}}
\end{eqnarray}
\begin{eqnarray}\label{eq:2}
 \hat{H}_{{\rm atom}}^{\alpha\beta} = \sum_{j=1,2}\left [ -\frac{\hbar^2}{2m_{a}}\nabla_j^2   + V_j^{\rm{trap}}({\bf r}_j)  \right ] + V_{\rm{\alpha\beta}}(|{\bf{r}_1} - \bf{r}_2 |)
\end{eqnarray}
is the part that depends only on the atomic COM DOF with 
\begin{eqnarray}\label{eq:3}
V_j^{\rm{trap}}({\bf r}_j) = \frac{1}{2}m_a \left [ \omega_{a\rho}^2 \left \{ (x_j - x_{j0})^2 + (y_j - y_{j0})^2 \right \} + \omega_{a z}^2 ( z_j -z_{j 0 })^2 \right ]    
\end{eqnarray}
being the harmonic trapping potential  of the $j$th atom and $ V_{\rm{\alpha\beta}}(|{\bf{r}_1} - \bf{r}_2 |)$ the interatomic potential. Here $\omega_{a\rho}$
and $\omega_{a z}$ denote the radial and axial trapping frequencies of the atom, respectively; $(x_{j0}, y_{j0}, z_{j0})$
is the coordinate of the $j$the atom's trap center. The axial and radial length scales associated with the atoms' trapping motion are $a_{z}=\frac{\hbar}{m_{a}\omega_{z}}$ and $a_{\rho}=\frac{\hbar}{m_{a}\omega_{a\rho}}$. The Hamiltonian $H_{\rm ion-atom}$ describing the motion of the ion in the presence of the ion-atom interactions is given by 
\begin{eqnarray}\label{eq:4}
 \hat{H}_{\rm ion-atom} &=& -\frac{\hbar^2}{2m_i}\nabla_i^2 +  V_{\rm ion-atom};\\\label{eq:5}
  V_{\rm ion-atom} &=& \frac{1}{2} m_{i}\left [ \omega_{i\rho}^2 \rho_i^2 +  \omega_{i z}^2 z_i^2 \right ]   
 -  \frac{C_4^{\alpha}}{|{\bf r}_i-{\bf r}_{1}|^4}-  \frac{C_4^{\beta}}{|{\bf r}_i-{\bf r}_{2}|^4}
\end{eqnarray}
where ${\rho}_i^2 = x_i^2 + y_i^2 $ with $(x_i, y_i, z_i)$ being the position of the ion. Here $\omega_{i r}$ and $\omega_{i z}$ are the radial and axial trapping frequencies of the ion, respectively,  and $C_{4}^{\alpha(\beta)}$ is the long range coefficient of interaction between ion and the atom $A_1$($A_2$).

\subsection{Born-Oppenheimer approximation}
The Born-Oppenheimer approximation we apply here to solve the problem relies on the smallness of the ratio of the speed of atom's  motion to the speed of the ion's motion. Since we consider that  the ion-trap as well as the two atom traps are cylindrical, one may define an average trapping  frequency  of the ion  $\bar{\omega}_{i} = [\omega_{i\rho}^2 \omega_{i z} ]^{1/3}$, and similarly an average atomic trapping frequency  $\bar{\omega}_{a} = [\omega_{a\rho}^2 \omega_{a z} ]^{1/3}$. Correspondingly, one can define average time and length scales of the ionic  motion by 
$T_i = 2 \pi/\bar{\omega}_i$ and $L_i = \sqrt{\hbar/(m_i \bar{\omega}_i)}$, respectively; and those of the atomic motion by $T_a = 2 \pi/\bar{\omega}_a$ and $L_a = \sqrt{\hbar/(m_a \bar{\omega}_a)}$. Thus one can define the ratio $\eta = {L_a T_i}/{L_i T_a}$  of the average speed of atom to the average speed  of ion by 
\bea \label{eq:5}
\eta = \sqrt{\frac{m_i}{m_a}} \left (\frac{\bar{\omega}_a}{\bar{\omega}_i}  \right )^{1/2}
\eea
Since, typically $\bar{\omega}_a/\bar{\omega}_i \sim 10^{-2}$, for $m_i \le m_a $, $\eta < 1$. So, one can safely apply BO approximation to separate out the atomic and ionic motion as a first approximation to obtain adiabatic solution of the problem. Later non-adiabatic effects may be included to improve the solutions. Note that the criterion $\eta < 1$ here is different from the criterion for BO approximation in molecular physics where one can separate out the much faster motion of electrons from the nuclear motion of a molecule - in molecular physics the counterpart of $\eta$ is the square root of the ratio between the electron's mass to the nuclear mass. 

 We diagonalize the Hamiltonian $\hat{H}_{\rm{ion-atom}}$ to obtain ion-mediated adiabatic BO potentials. Let $\psi_{\mu}^{\alpha\beta} (\bf{r_i}; \bf{r_1}, \bf{r_2})$ represent an eigenfunction with corresponding eigenvalue $ V_{\mu}^{{\alpha\beta}}( \bf{r_1}, \bf{r_2})$, where $\mu$ denotes a collection of quantum numbers that characterize an eigenstate. Both the eigenfunction and  eigenvalue parametrically depend on the atomic position coordinates $\bf{r_1}$ and   $\bf{r_2}$ which are slowly varying in time. 

\subsection{Adiabatic ionic and atomic states}
In order to derive analytical solution, we approximate the potential  $V_{\rm ion-atom}$ 
 of Eq.\eqref{eq:5} as 
 \begin{eqnarray}\label{eq:7}
    V_{\rm ion-atom} &\approx& \frac{1}{2} m_{i} \omega_{i\rho}^2 \tilde{\rho}_{i}^2 + \frac{1}{2} m_{i} \omega_{i z}^2 \tilde{z}_{i}^2 - \frac{1}{2} m_{i}\omega_{i\rho}^2 \rho_0^{2} \nonumber\\ &-&  \frac{1}{2} m_{i} \omega_{i z}^2 \zeta_0^{2} - \left(\frac{C_{4}^{\alpha}}{r_{1}^4} + \frac{C_{4}^{\beta}}{r_{2}^4}\right)-\frac{C^{\alpha\beta}_{6}}{\left(|\bf r_{1} -\bf r_{2}| \right)^{6}}
\end{eqnarray}
where $\tilde{\rho}_{i}^{2} = \left(x_{i}-x_0\right)^{2} +\left(y_{i}-y_{0}\right)^{2}$, $\tilde{z}_{i}^{2}=\left(z_{i}-\zeta_{0}\right)^{2}$, $\rho_{0}^{2}=x_{0}^{2}+y_{0}^{2}$ with $x_0 = \frac{4}{m_i \omega_{i\rho}^2} \left( \frac{C_{4}^{\alpha}x_{1}}{r_{1}^6} +\frac{C_{4}^{\beta}x_{2}}{r_{2}^6}\right)$, $y_0 = \frac{4}{m_i \omega_{i\rho}^2} \left( \frac{C_{4}^{\alpha}y_{1}}{r_{1}^6} +\frac{C_{4}^{\alpha}y_{2}}{r_{2}^6}\right)$ and 
$\zeta_0 = \frac{4}{m_i \omega_{iz}^2} \left( \frac{C_{4}^{\alpha}z_{1}}{r_{1}^6} +\frac{C_{4}^{\alpha}z_{2}}{r_{2}^6}\right)$. $C_{6}^{\alpha\beta}$ is the Van der Waals coefficient between two atoms in $|\alpha\rangle$ and $|\beta\rangle$ internal states. In writing Eq.\eqref{eq:7}, we have expanded $V_{\rm ion-atom}$ up to first order in $\frac{r_{i}}{r_{j}}$, since $\frac{r_{i}}{r_{j}} <<1$. To calculate the ionic state we solve eigenvalue equation
\begin{eqnarray}\label{eq:8}
    \hat{H}_{\rm ion-atom}\psi_{\mu}\left(\bf r_{i}, \bf r_{1}, \bf r_{2}\right)=V_{\mu}^{\alpha\beta}\psi_{\mu}\left(\bf r_{i}, \bf r_{1}, \bf r_{2}\right)
\end{eqnarray}
The eigenvalue $V_{\mu}^{\alpha\beta}$ is given by 
\begin{eqnarray}\label{eq:9}
    V_{\mu}^{\alpha\beta}\left(\bf {r}_1, \bf {r}_2\right) &=& \left(2n_{\rho_{i}} + \left|m_{i}\right| + 1\right)\hbar \omega_{i\rho} + \left(n_{z_{i}} + \frac{1}{2}\right)\hbar \omega_{iz} + \frac{1}{2} m_{i}\omega_{i\rho}^2 \rho_0^{2} \left(r_1, r_2\right) \nonumber\\ &+&  \frac{1}{2} m_{i} \omega_{i z}^2 \zeta_0^{2} \left(r_1, r_2\right) - \left(\frac{C_{4}^{\alpha}}{r_{1}^4} + \frac{C_{4}^{\beta}}{r_{2}^4}\right)-\frac{C^{\gamma_{12}}_{6}}{\left(|\bf r_{1} -\bf r_{2}| \right)^{6}}                 
\end{eqnarray}
where $\mu \equiv \left(n_{\rho_{i}}, m_{i}, n_{z_{i}}\right)$ with $n_{\rho_{i}}, m_{i}, n_{z_{i}}$ being the radial, azimuthal and axial quantum numbers respectively, $n_{\rho}=0,1,2...$, $m=0, \pm 1...$ and $n_{z}=0,1,2...$. The eigen state $\psi_{\mu}$ is given by
\begin{eqnarray}\label{eq:10}
\psi_{\mu}=\frac{1}{\sqrt{2\pi}}\psi_{n_{z_{i}}}\left(\tilde{z}_{i}\right)\psi_{n_{\rho_{i}}}\left(\tilde{\rho}_{i}\right)e^{im_{i}\phi_{i}}
\end{eqnarray}
where $\psi_{n_{z_{i}}}\left(\tilde{z}_{i}\right)$ represents 1D harmonic oscillator state in the $z$-axis, $\psi_{n_{\rho_{i}}}\left(\tilde{\rho}_{i}\right)$ represents the radial wave function and $\phi$ is the azimuthal angle of the ion. Thus adiabatic BO state $\phi_{\mu}\left(\bf r_{1}, \bf r_{2}\right)$ can be obtained by solving the eigenvalue equation
\begin{eqnarray}\label{eq:11}
    \left[-\sum_{j=1,2}\frac{\hbar^{2}}{2m_{a}}\nabla_{j}^{2}+U_{\mu}^{\alpha\beta}\right]\phi_{\mu}\left(\bf r_{1}, \bf r_{2}\right)=E_{\mu}\phi_{\mu}\left(\bf r_{1}, \bf r_{2}\right)
\end{eqnarray}

where
\begin{eqnarray}\label{eq:12}
    U_{\mu}^{\alpha\beta}\left(\bf {r}_1, \bf {r}_2\right) &=& \frac{1}{2}m_{a}\omega_{az}^{2}\left((z_{1}+z_0)^2+(z_{2}-z_0)^2\right)+\frac{1}{2}m_{a}\omega_{a\rho}^{2}\sum_{j=1}^2\rho_{j}^{2} +V_{\mu}^{\alpha\beta}\left({\bf r_1,r_2}\right)
\end{eqnarray}

In order to calculate $\phi_{\mu}\left(\bf r_{1}, \bf r_{2}\right)$ and $E_{\mu}$ perturbatively, we make some approximation in $U_{\mu}^{\alpha\beta}\left(\bf r_{1}, \bf r_{2}\right)$.
The positions of the centers of the traps for atom $A_{1}$ and $A_{2}$ are set at $\left(0, 0, z_{0}\right)$ and $\left(0, 0, -z_{0}\right)$ respectively. By introducing the variables $z_{1}'=z_{1} - z_{0}$ and $z_{2}'=z_{2}+z_{0}$, Eq.\eqref{eq:11} can be approximated as
\begin{eqnarray}\label{eq:13}
     U_{\mu}^{\alpha\beta}\left({\bf r_{1}, r_{2}}\right) &\approx& E_{\mu}^{(0)} +\frac{1}{2}m_{a}\omega_{az}^{2}\left(z_{1}'^{2} + z_{2}'^{2}\right) +\frac{1}{2}m_{a}\sum_{j=1,2}\left(\omega_{a\rho}^{2} -\frac{A_{12}^{\gamma_{j}}}{\left(z_{1}'+z_{0}\right)^{12}}-\frac{A_{6}^{\gamma_{j}}}{\left(z_{1}'+z_{0}\right)^{6}}\right)\rho_{j}^{2} \nonumber\\ &-&  \frac{{m_a}A_{12}^{\alpha\beta}\rho_{1}\rho_{2}\cos\left(\phi_{1}-\phi_{2}\right)}{\left(z_{1}'+z_{0}\right)^{6}\left(z_{2}'-z_{0}\right)^{6}} - \frac{m_{a}A_{10}^{\alpha\beta}}{\left(z_{1}'+z_{0}\right)^{5}\left(z_{2}'+z_{0}\right)^{5}} -\frac{m_{a}A_{10}^{\gamma_{1}}}{2\left(z_{1}'+z_{0}\right)^{10}} -
     \frac{m_{a}A_{4}^{\gamma_{1}}}{2\left(z_{1}'+z_{0}\right)^{4}}\nonumber \\ &-&
     \frac{m_{a}A_{10}^{\gamma_{2}}}{2\left(z_{2}'-z_{0}\right)^{10}} -\frac{m_{a}A_{4}^{\gamma_{2}}}{2\left(z_{2}'-z_{0}\right)^{4}}-\frac{C_{6}^{\alpha\beta}}{\left\{(\rho_1-\rho_2)^{2}+\left(z'_1-z'_2+2z_{0}\right)^{2}\right\}^3}
\end{eqnarray} 
where $E_{\mu}^{(0)}=\left(2n_{\rho_{i}} + \left|m_{i}\right| + 1\right)\hbar \omega_{i\rho} + \left(n_{z_{i}} + \frac{1}{2}\right)\hbar \omega_{iz}$, $A_{12}^{\gamma_{j}}=\frac{16\left(C_{4}^{\gamma_{j}}\right)^{2}}{m_{a}m_{i}\omega_{i\rho}^{2}}$, $A_{12}^{\alpha\beta}=\frac{16C_{4}^{\alpha}C_{4}^{\beta}}{m_{a}m_{i}\omega_{i\rho}^{2}}$, $A_{6}^{\gamma_{j}} = \frac{4C_{4}^{\gamma_{j}}}{m_{a}}$, $A_{10}^{\gamma_{j}}=\frac{16\left(C_{4}^{\gamma_{j}}\right)^{2}}{m_{a}m_{i}\omega_{iz}^{2}}$, $A_{10}^{\alpha\beta}=\frac{16C_{4}^{\alpha}C_{4}^{\beta}}{m_{a}m_{i}\omega_{iz}^{2}}$, $A_{4}^{\gamma_{j}}=\frac{2C_{4}^{\gamma_{j}}}{m_{a}}$
with $\gamma_{1}=\alpha$ and $\gamma_{2}=\beta$. Here $\phi_{1}$ and $\phi_{2}$ are the azimuthal angles of the atoms $A_{1}$ and $A_{2}$, respectively. As $2z_{0}>>a_{\rho}, a_{z}$, we further approximate the Eq.\eqref{eq:12} up to the quadratic terms in atomic coordinates. The 
potential takes the form
\begin{eqnarray}\label{eq:14}
    U_{\mu}^{\alpha\beta}\left(\bf {r}_1, \bf {r}_2\right)=T_{1}+T_{2}+\bar E_{\mu}^{(0)}
\end{eqnarray}
where 
\begin{eqnarray}
    T_{1}&=&\frac{1}{2}m_{a}\sum_{j=1,2}\left[ \omega_{a\rho}^{2}-\frac{A_{6}^{\gamma_{j}}}{z_{0}^{6}}-\frac{A_{12}^{\gamma_{j}}}{z_{0}^{12}}+\frac{3C_6^{\alpha\beta}}{128m_{a}z_0^8}\right]\rho_{j}^{2} \nonumber\\
    &+&\frac{1}{2}m_{a}\sum_{j=1,2}\left[ \omega_{az}^{2}-\frac{10A^{\gamma_j}_{4}}{z_0^{6}}- \frac{55A^{\gamma_j}_{10}}{z_0^{12}}-\frac{15A_{10}^{\alpha\beta}}{z_0^{12}}-\frac{21C_6^{\alpha\beta}}{128m_az_0^8} \right]z_j'^2\nonumber\\
    T_{2}&=& - m_{a}\left(\frac{A_{12}^{\alpha\beta}}{z_{0}^{12}} + \frac{3C_6^{\alpha\beta}}{128m_{a}z_0^8}\right)\cos\left(\phi_{1}-\phi_{2}\right)\rho_{1}\rho_{2} + m_{a}\left(\frac{25A_{10}^{\alpha\beta}}{z_0^{12}}+\frac{21C_6^{\alpha\beta}}{128m_{a}z_0^8}\right)z'_1z'_2 \nonumber \\ &+& m_a{\Bigg[}{\Bigg(}\frac{5A_{10}^{\alpha\beta}}{z_0^{11}}+\frac{5A_{10}^{\gamma_{1}}}{z_0^{11}}+\frac{2A_4^{\gamma_1}}{z_0^5}-\frac{3C_6^{\alpha\beta}}{64m_az_0^7}{\Bigg)}z'_1 - {\Bigg(}\frac{5A_{10}^{\alpha\beta}}{z_0^{11}}+\frac{5A_{10}^{\gamma_{2}}}{z_0^{11}}+\frac{2A_4^{\gamma_2}}{z_0^5}-\frac{3C_6^{\alpha\beta}}{64m_az_0^7}{\Bigg)}z'_2 {\Bigg]}\nonumber
\end{eqnarray}
and $\bar E_\mu^{(0)}=E_\mu^{(0)}-m_a\left(\frac{A_4^{\gamma_1}}{2z_0^4}+\frac{A_4^{\gamma_2}}{2z_0^4}+\frac{A_{10}^{\gamma_1}}{2z_0^{10}}+\frac{A_{10}^{\gamma_2}}{2z_0^{10}}+\frac{A_{10}^{\alpha\beta}}{2z_0^{10}}\right)-\frac{C_6^{\alpha\beta}}{64z_0^6}$. The first term $T_{1}$ describes uncoupled harmonic motion of the two atoms with modified radial and axial frequencies 
\begin{eqnarray}
\label{eq:15}
\bar\omega_{a\rho}^{j} &=& \sqrt{\omega_{a\rho}^{2}-\frac{A_{6}^{\gamma_{j}}}{z_{0}^{6}}-\frac{A_{12}^{\gamma_{j}}}{z_{0}^{12}}+\frac{3C_6^{\alpha\beta}}{128m_{a}z_0^8}}\\ 
\bar\omega_{az}^j &=& \sqrt{\omega_{az}^{2}-\frac{10A^{\gamma_j}_{4}}{z_0^{6}}- \frac{55A^{\gamma_j}_{10}}{z_0^{12}}-\frac{15A_{10}^{\alpha\beta}}{z_0^{12}}-\frac{21C_6^{\alpha\beta}}{128m_az_0^8}}
\label{eq:16}
\end{eqnarray}
The term $T_{2}$ will couple even radial harmonic function with odd ones and vice versa of each atom while the even azimuthal angular function will be coupled to odd ones and vice versa. This means the third term will lead to transitions $|n_{\rho_{1}},n_{\rho_{2}}, m_{1}, m_{2}\rangle \rightarrow |n_{\rho_{1}}\pm 1,n_{\rho_{2}}\pm 1, m_{1}\pm 1, m_{2}\pm 1\rangle$. For the stability of the system, we must have $\bar{\omega}_{a\rho}^{j}>0$ and $\bar{\omega}_{az}^{j}>0$, implying that there exists a minimum separation $2z_{0}$ below which the system is unstable due to direct collision between the ion and the atoms.

\subsection{Phonon modes}
 To calculate the phonon modes of the atoms, we rewrite the potential of Eq.\eqref{eq:14} in terms of scaled relative and centre-of-mass (COM) coordinates
    ${\bf r}=({\bf r_{1} - r_{2}})/\sqrt{2}$ and
    ${\bf R}= ({\bf r_{1} + \bf r_{2}})/\sqrt{2}$, respectively.
    In cartesian coordinates Eq.\eqref{eq:14} can be written as
    \begin{eqnarray}\label{eq:17}
    U_{\mu}^{\alpha\beta} &=&\frac{1}{2}m_a\left[\left\{\left(\omega_{a\rho}'\right)^{2}+\left(\omega_{xy}\right)^{2}\right\}\left(x^{2}+y^{2}\right)+\left\{\left(\omega_{az}'\right)^{2}-\left(\omega_{zz}\right)^{2} \right\}z^{2}\right] \nonumber \\ &+& \frac{1}{2}m_a\left[\left\{\left(\omega_{a\rho}'\right)^{2}-\left(\omega_{xy}\right)^{2}\right\}\left(X^{2}+Y^{2}\right)+\left\{\left(\omega_{az}'\right)^{2} +\left(\omega_{zz}\right)^{2}\right\}Z^{2}\right]\nonumber \\ &+& \frac{1}{2}m_{a}\left\{\left(\bar\omega_{a\rho}^\alpha\right)^2 - \left(\bar\omega_{a\rho}^\beta\right)^2\right\}\left(xX+yY\right)+\frac{1}{2}m_{a}\left\{\left(\bar\omega_{az}^\alpha\right)^2 - \left(\bar\omega_{az}^\beta\right)^2\right\}zZ\nonumber \\ &+& m_{a}z_{0}\frac{\left(\Omega_{1}^{2} + \Omega_{2}^{2} \right)}{\sqrt{2}}z +  m_{a}z_{0}\frac{\left(\Omega_{1}^{2} - \Omega_{2}^{2} \right)}{\sqrt{2}}Z \nonumber \\ &+& \left(n_{x_{i}} + \frac{1}{2}\right)\hbar \omega_{ix} + \left(n_{y_{i}} + \frac{1}{2}\right)\hbar \omega_{iy} + \left(n_{z_{i}} + \frac{1}{2}\right)\hbar \omega_{iz}
    \label{eq.17}
\end{eqnarray}
Here we introduce some frequency scales as
\begin{eqnarray}
\omega_{a(\rho, z)}'^{2}&=&\frac{\left(\bar\omega_{a(\rho, z)}^\alpha\right)^2 + \left(\bar\omega_{a(\rho, z)}^\beta\right)^2}{2},
    \omega_{xy}^{2} = \left(\frac{A_{12}^{\gamma_{12}}}{z_{0}^{12}} + \frac{3C_6^{\alpha\beta}}{128m_{a}z_0^8}\right),  \nonumber \\
    \omega_{zz}^{2} &=& \left(\frac{25A_{10}^{\gamma_{12}}}{z_0^{12}}+\frac{21C_6^{\alpha\beta}}{128m_{a}z_0^8}\right), 
    \Omega_{j}^{2} = \frac{5A_{10}^{\alpha\beta}}{z_0^{12}}+\frac{5A_{10}^{\gamma_{j}}}{z_0^{12}}+\frac{2A_4^{\gamma_j}}{z_0^6}-\frac{3C_6^{\alpha\beta}}{64m_az_0^8} \nonumber
\end{eqnarray}
and $\{x, y, z\} = \{\frac{x_{1}-x_{2}}{\sqrt{2}},\frac{y_{1}-y_{2}}{\sqrt{2}}, \frac{z_{1}-z_{2}}{\sqrt{2}}\}$ and $\{X, Y, Z\} = \{\frac{x_{1}+x_{2}}{\sqrt{2}}, \frac{y_{1}+y_{2}}{\sqrt{2}}, \frac{z_{1}+z_{2}}{\sqrt{2}}\}$ are relative and COM certesian coordinates, respectively.
The transverse and axial frequency  of the harmonic oscillators are modified. From Eq.(\ref{eq:17}) it is evident that the transverse and longitudinal phonon modes are separable. Also if the two atoms are prepared in the same internal state, i.e., $\alpha = \beta$, then the relative or stretched and COM modes are separable. Phonon modes are calculated by diagonalizing the Heissian matrix
\begin{eqnarray}
\mathcal{A}^{\alpha\beta}_\mu = \frac{1}{m_a}\begin{pmatrix}
    \frac{\partial^2{U_\mu^{\alpha\beta}}}{\partial{u^2}}      &\frac{\partial^2{U_\mu^{\alpha\beta}}}{\partial{u\partial{v}}}\\ 
  \frac{\partial^2{U_\mu^{\alpha\beta}}}{\partial{v\partial{u}}}      &\frac{\partial^2{U_\mu^{\alpha\beta}}}{\partial{v^2}}
\end{pmatrix}\nonumber
\end{eqnarray}
where for transverse modes $\{u,v\}\equiv\{x,X\}$ or $\{y,Y\}$ and for axial modes $\{u,v\}\equiv\{z,Z\}$. 

\subsection{Nonadiabatic effects}
By incorporating non-adiabatic effects, a general wave function $\Psi(\bf{r}_i, \bf{r}_1(t), \bf{r}_2(t),t)$ of the entire system may be expanded in terms of the ionic eigenfunctions $\psi_{\mu} (\bf{r_i}; \bf{r_1}(t), \bf{r_2}(t))$
 \begin{eqnarray}\label{Eq:18}
    \Psi(\bf{r}_i, \bf{r}_1(t), \bf{r}_2(t),t ) = \sum_{\mu} \tilde{\phi}_{\mu} (\bf{r}_1(t), \bf{r}_2(t),t)  \psi_{\mu} (\bf{r_i}; \bf{r_1}(t), \bf{r_2}(t))
 \end{eqnarray}
where $ \tilde{\phi}_{\mu} (\bf{r}_1 (t), \bf{r}_2 (t), t)$ denotes two-atom wave function with non-adiabatic effects.
Substituting Eq.\eqref{Eq:18} in the Schr\"{o}dinger equation $\hat{H}\Psi=i\hbar\frac{\partial \Psi}{\partial t}$, we have
\begin{eqnarray}\label{eq:19}
    \left[-\frac{\hbar^{2}}{2m_{a}}\sum_{j=1,2}\nabla_{j}^2+U_{\nu}^{\alpha\beta}\left(\bf {r}_1, \bf {r}_2\right)\right]\tilde{\phi}_{\nu}+\frac{i\hbar}{m_{a}}\sum_{\mu}\sum_{j}\bf {A}_{\nu\mu}^{j}\cdot\nabla_{r}\tilde{\phi}_{\mu}\\ \nonumber + \frac{1}{2}\sum_{\mu}\sum_{j}\left(\bf {A}_{\nu\mu}^{j}\right)^{2}\tilde{\phi}_{\mu}&=&i\hbar\frac{\partial \tilde{\phi}_{\nu}}{\partial t}
\end{eqnarray}
\begin{eqnarray}\label{eq:20}\hspace{-1.75in}{\rm where}
\hspace{0.5in}
{\bf A}_{\mu\nu}^{j}({\bf r_1,r_2}) = i\hbar\int\psi_\nu({\bf r_i;r_1,r_2})\nabla_{\bf r_j}\psi_\mu({\bf r_i;r_1,r_2})d^3r_i
\end{eqnarray}
is a vector potential or a Gauge field. If one considers non-adiabatic effects under BO approximation, then  $\nu=\mu$. Equation(\ref{eq:19}) resembles to the Hamiltonian of a charged particle in a Gauge field. One can  make the transformation
\begin{eqnarray}\label{eq:21}
    \tilde{\phi}_\mu({\bf r_1(t),r_2(t)}, t)=e^{i\gamma_{\mu} (C)}\phi_\mu({\bf r_1(t),r_2(t)}, t)
\end{eqnarray} 
where
\begin{eqnarray}\label{eq:22}
    \gamma_{\mu} (C) = \oint_C \underline{\underline{\bf {A}}}_{\mu} \cdot d\underline{\underline{\bf {r}}}
\end{eqnarray}
 is the well-known geometric or Berry phase. Here $C$ refers to a closed path in $\underline{\underline{\bf {r}}} \equiv ({\bf r_1,r_2})$   space and
 $\underline{\underline{\bf {A}}}_{\mu}=\left(\bf {A}_{\mu}^{1}, \bf {A}_{\mu}^{2}\right)$. Thus  $\phi_\mu({\bf r_1(t),r_2(t)}, t)$ satisfies the equation 
 \begin{eqnarray}\label{eq:23}
     \left[-\frac{\hbar^{2}}{2m_{a}}\sum_{j=1,2}\nabla_{j}^2+U_{\nu}^{\alpha\beta}\left(\bf {r}_1, \bf {r}_2\right)\right]\phi_{\mu}({\bf r_1(t),r_2(t)}, t) = i\hbar\frac{\partial \phi_{\mu}({\bf r_1(t),r_2(t)}, t)}{\partial t}
 \end{eqnarray}
 The geometric phase $\gamma^j_{\mu} (C)$ will be non zero provided the closed path encloses a singularity or conical intersection \cite{Larson:Springer:2020}.

\section{Results and discussions} \label{Sec:3}
\label{Results and discussion}

\begin{figure}
    \includegraphics[height=1.5in,width=2.2in]{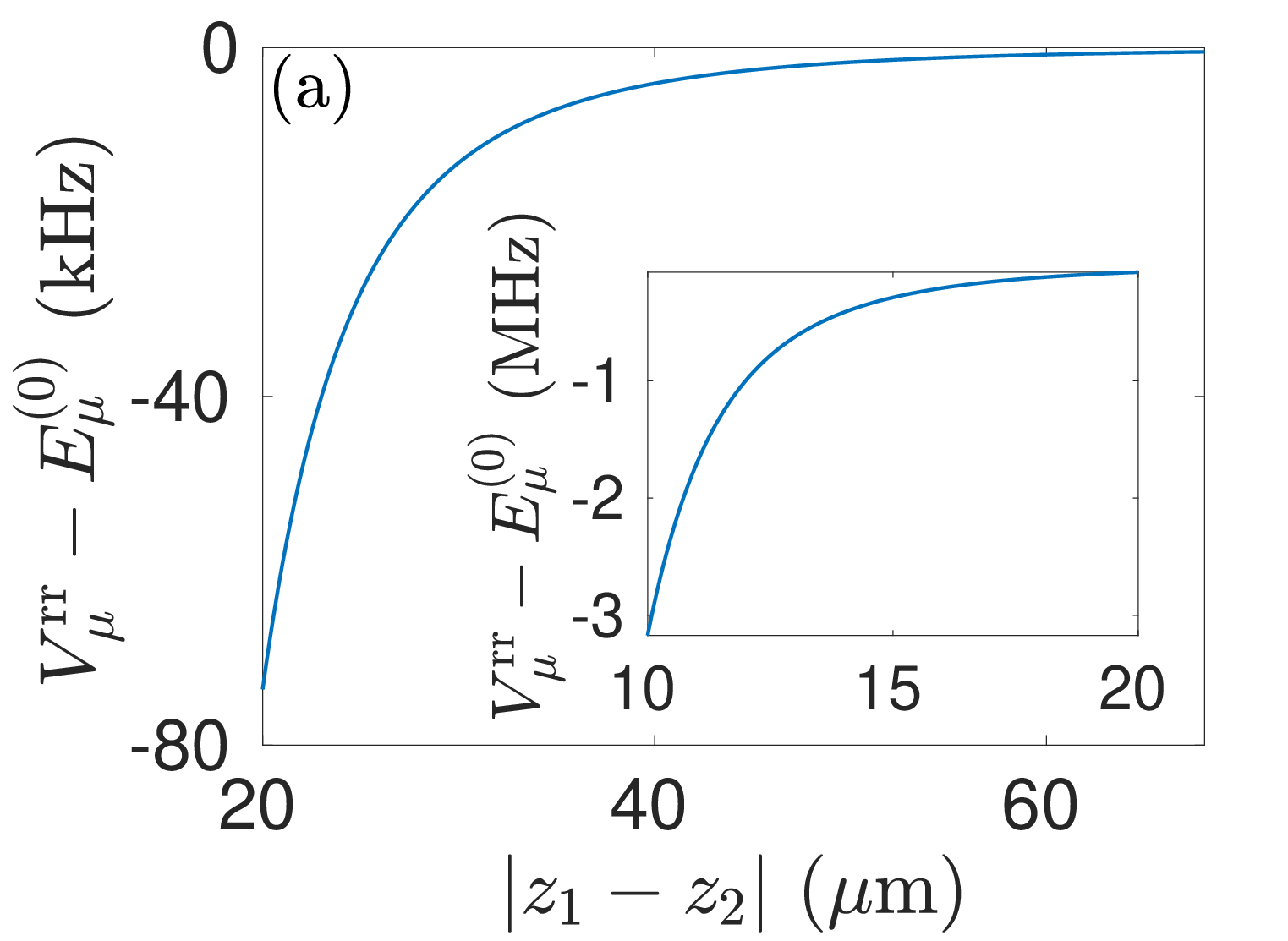}
    \hspace{-0.1in}
    \includegraphics[height=1.5in,width=2.2in]{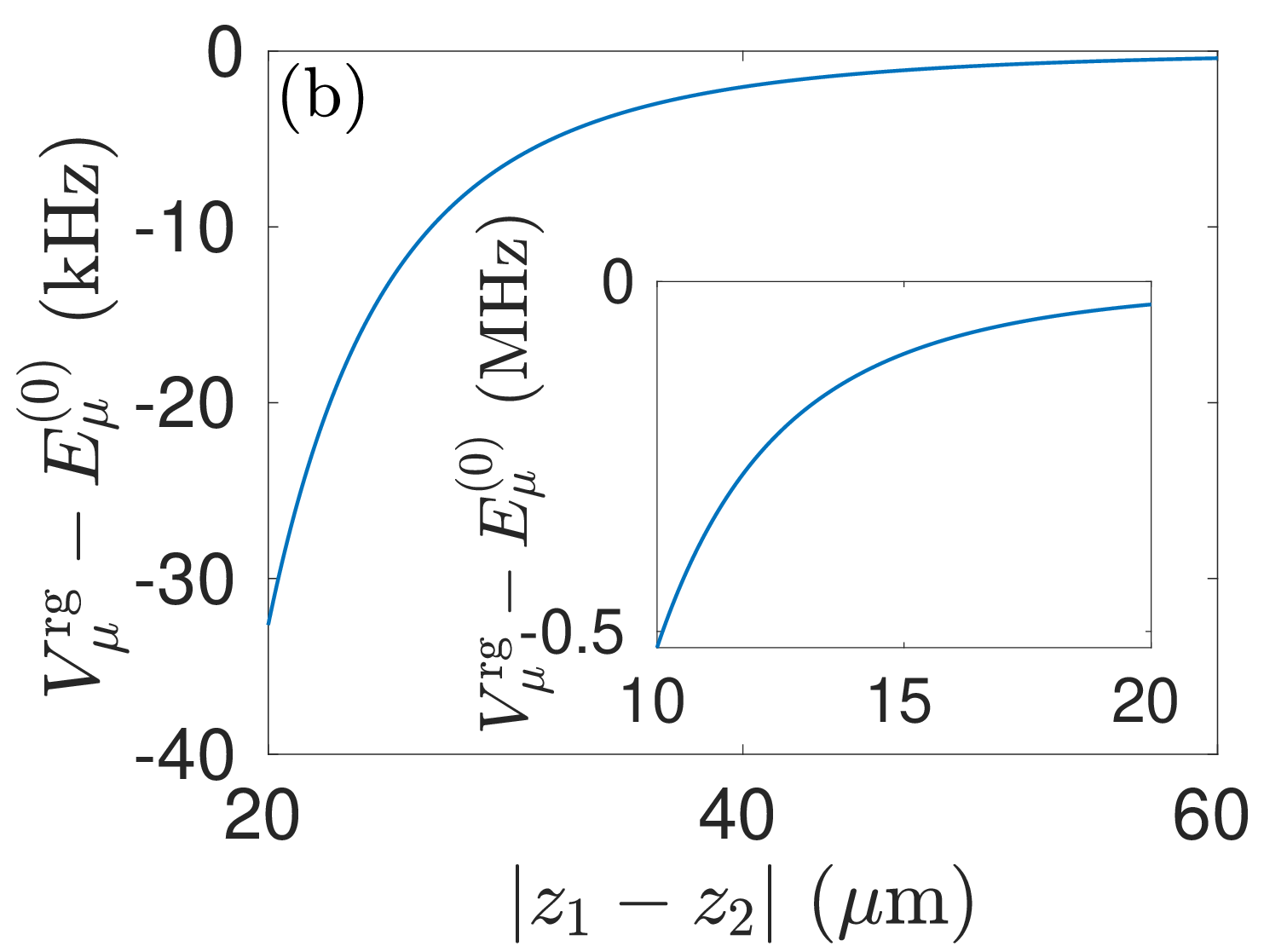}
    \hspace{-0.1in}
    \includegraphics[height=1.5in,width=2.2in]{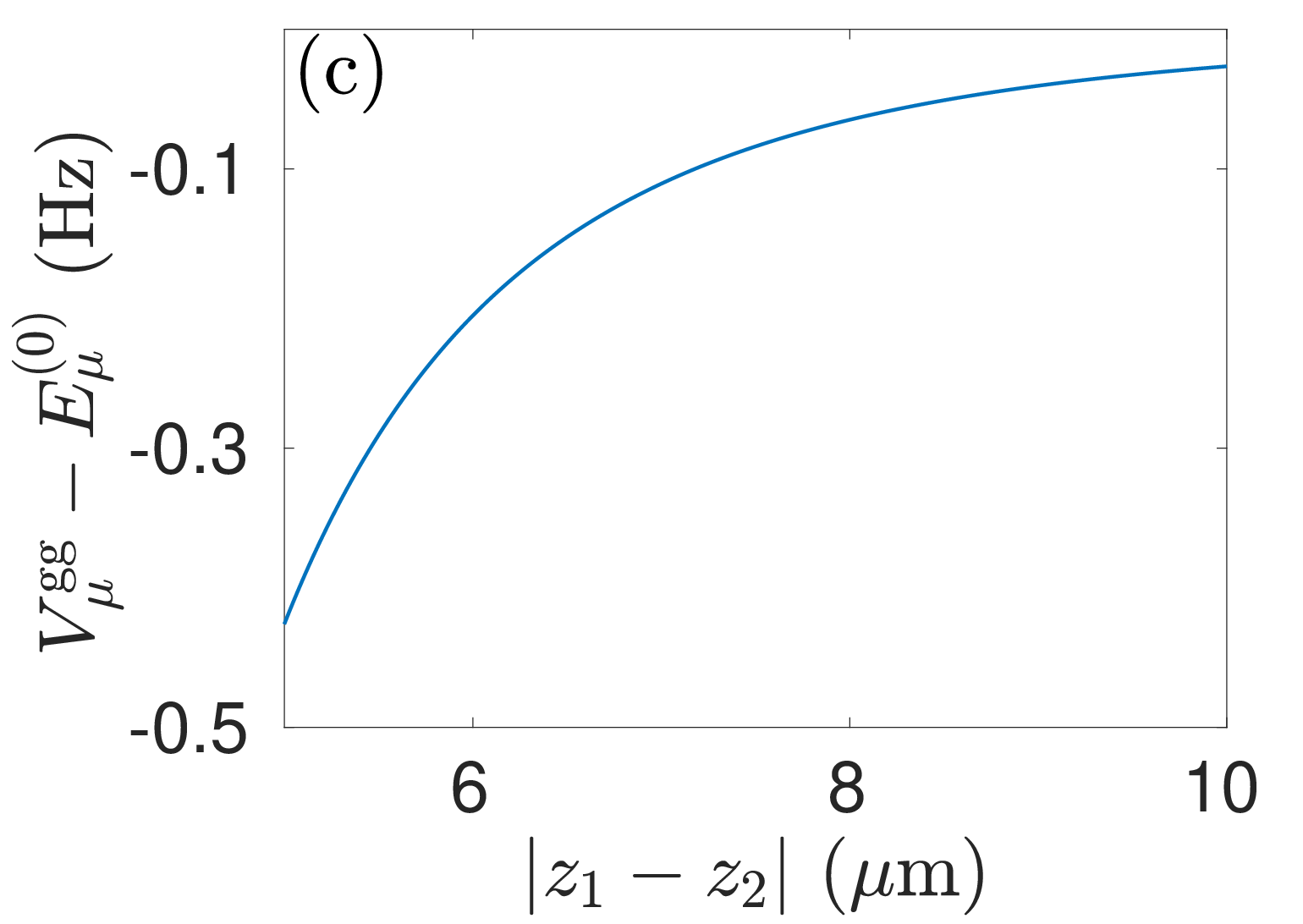}
  \caption{The lowest ($\mu \equiv \left(0,0,0\right)$)  Born-Oppenheimer potentials in quasi-1D case are plotted as a function of separation $z$ along the $z$-axis for the  {$^{87}$Rb+$^{40}$Ca$^{+}$+$^{87}$Rb} system.  (a) $V_{\mu}^{rr}$ (in kHz): both atoms are in the same Rydberg state ($30S-30S$), (b) $ V_{\mu}^{rg}$ (in kHz): one atom in the Rydberg state and other one is in the ground state ($30S-5S$) and (c) $V_{\mu}^{gg}$ (in Hz): both atoms are in the ground state ($5S-5S$). Insets of (a) and (b) show BO potential (in MHz) in the range 10-20 $\mu$m. }
  \label{fig 2}
\end{figure}
\begin{figure}
    \includegraphics[height=1.5in,width=6.5in]{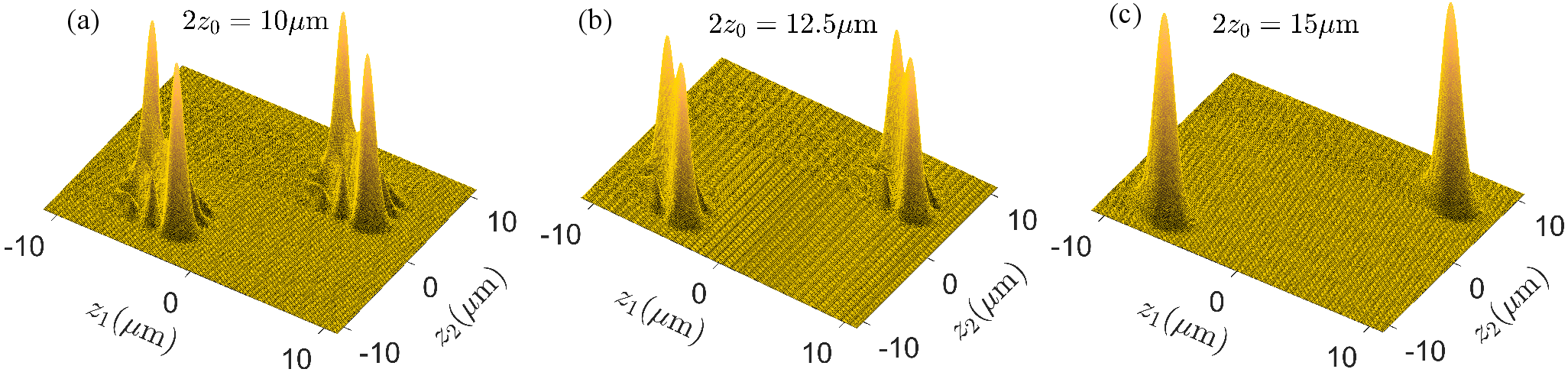}
    \caption{Axial probability densities of the pair of atoms as a function of individual atomic coordinates $z_1$ and $z_2$ for three different values of separation $2z_0$ between the atom-trap centers for the quasi-1D system as mentioned in Fig.\ref{fig 2}.}
  \label{fig 3}
\end{figure}
In the preceding section, we have derived some analytical results on the ion-mediated interaction between two Rydberg atoms, atomic phonon modes, and a synthetic Gauge structure and associated geometric phase. Here we present numerical results to further corroborate our analytical findings about phonon modes. We discuss briefly the consequences of the synthetic Gauge field and possible realization of a conical intersection in the system. For simplicity of calculations, we assume that the atom traps are quasi-one dimensional with $\omega_{a\rho}>>\omega_{az}$. Assuming that atoms are cooled to their radial ground state and the temperature is much smaller than $\hbar \omega_{a\rho}$, we use an ansatz for the atomic wave function as
\begin{eqnarray}\label{eq:24}
    \phi_{\mu}\left(\bf r_{1}, r_{2}\right) = \chi_{00}^{(1)}\left(\rho_{1}\right)\chi_{00}^{(2)}\left(\rho_{2}\right)\phi_{\mu}\left(z_{1}, z_{2}\right)
\end{eqnarray}
where $\chi_{00}^{(n)}\left(\rho_{n}\right)$ refers to the radial motional ground state of $n$th atom. By integrating over the radial ground states of the two atoms, we obtain an effective BO potential.
\begin{figure}
\centering
    \includegraphics[height=2.4in,width=3.2in]{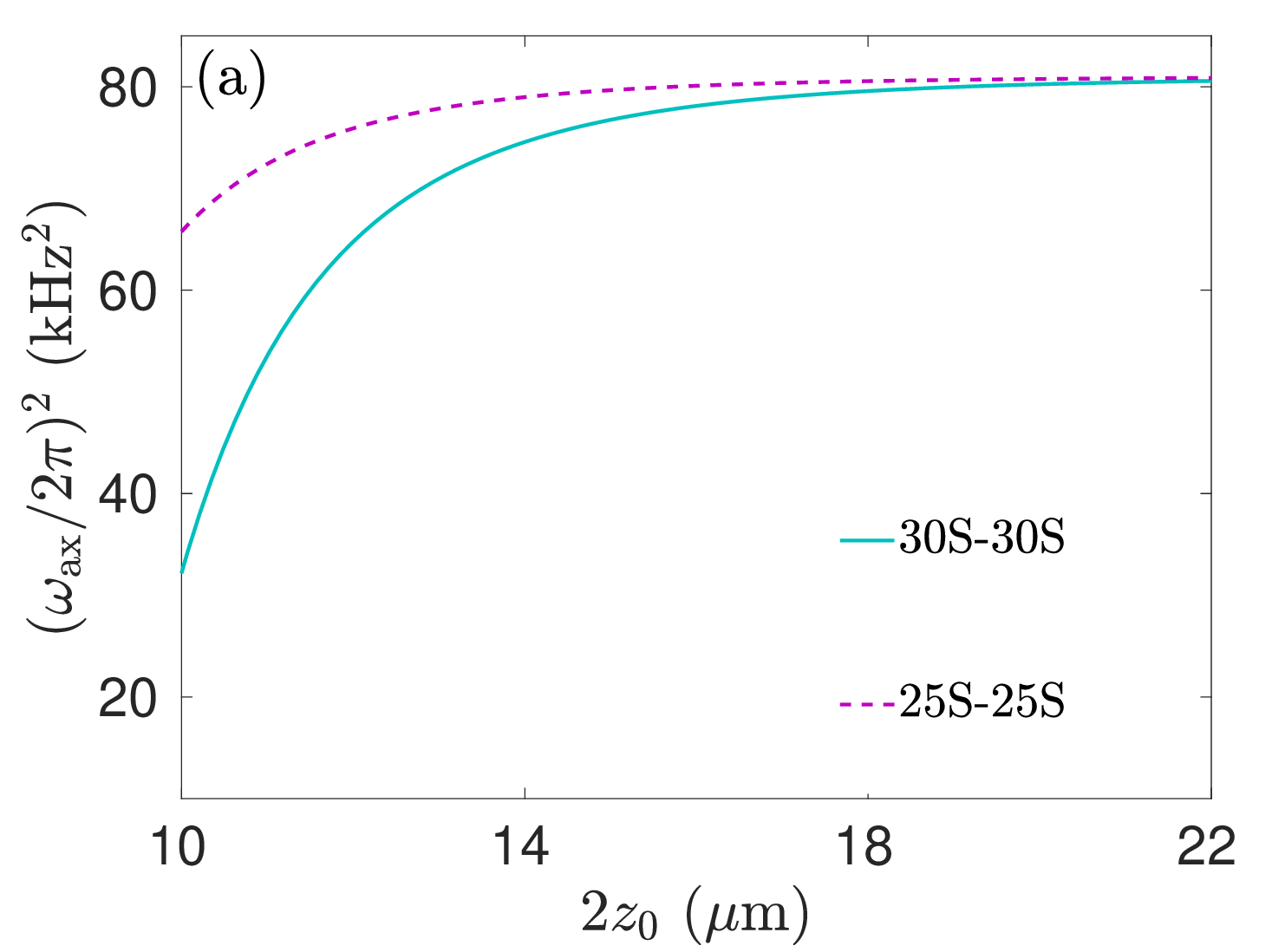}
    \hspace{-0.1in}
    \includegraphics[height=2.4in,width=3.5in]{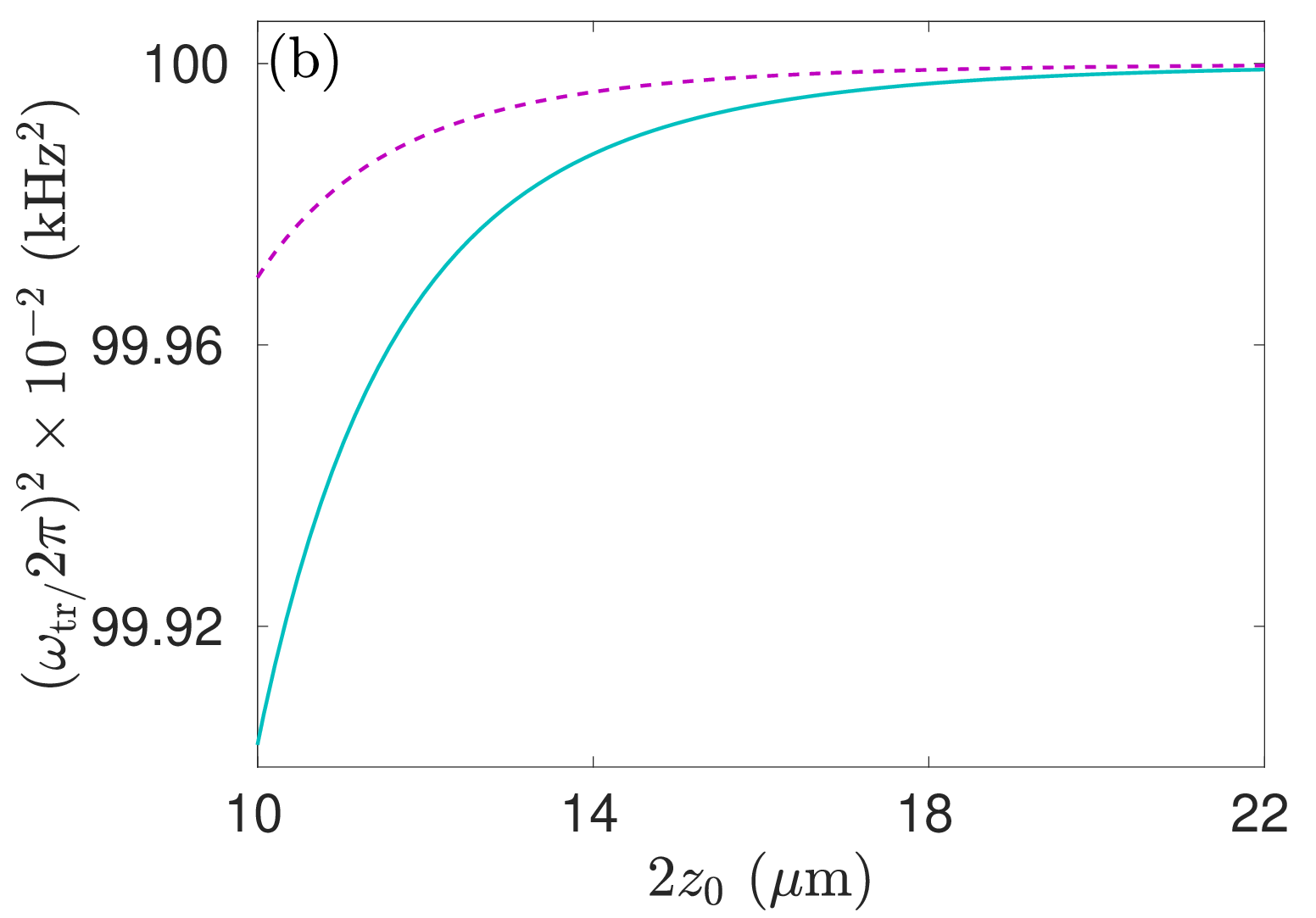}
  \caption{Variation of the square of axial (a) and transverse (b) phonon mode frequencies (in (kHz)$^{2}$) as a function of $2z_{0}$ for quasi-1D case of Fig.\ref{fig 2} when both atoms are in the state $30S-30S$ (solid line) and  $25S-25S$ (dashed line). Note that here st and COM modes are degenerate.}
  \label{fig 4}
\end{figure}
\begin{figure}
\centering
    \includegraphics[height=2.4in,width=3.2in]{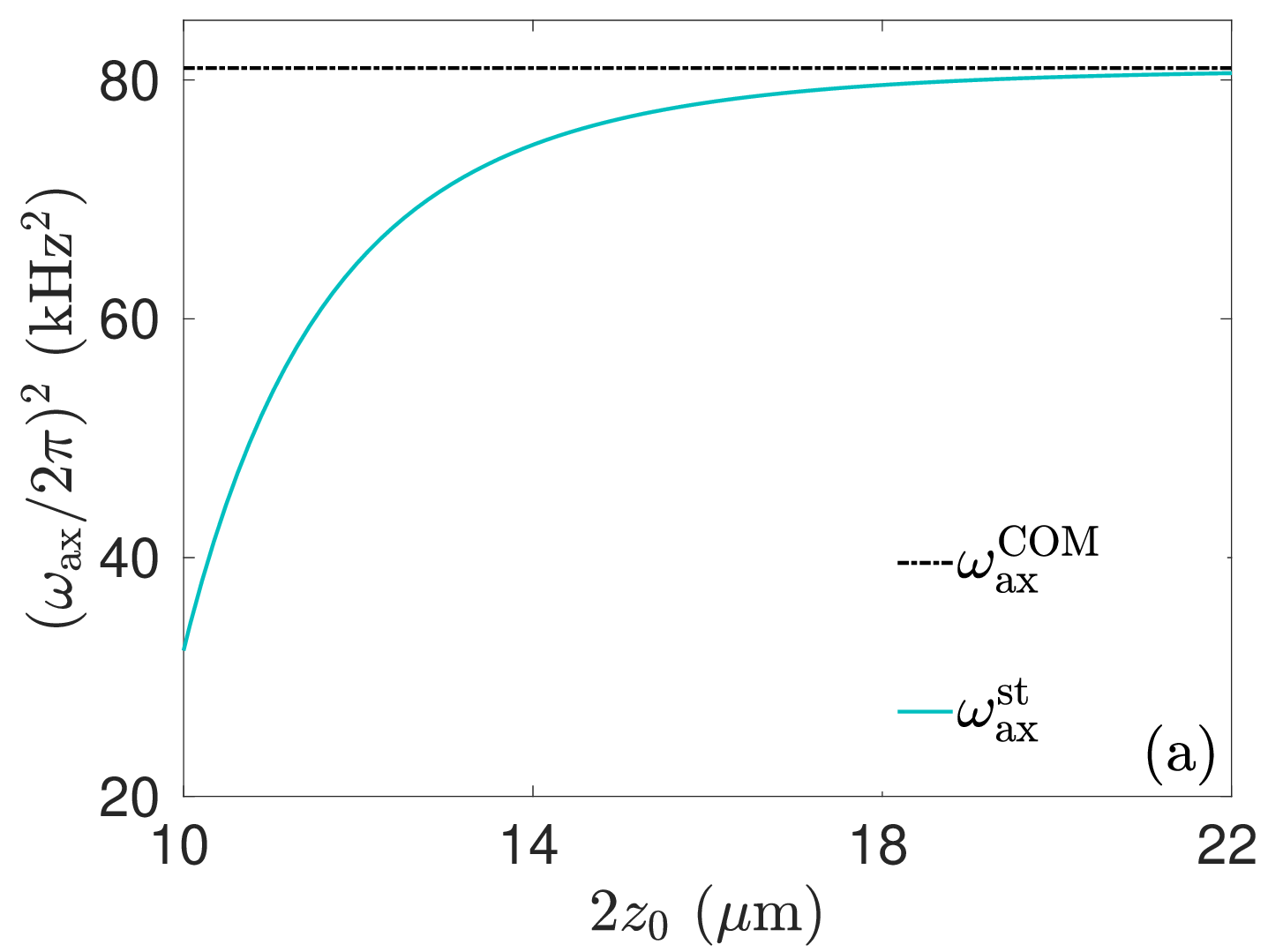}
    \hspace{0.11in}
    \includegraphics[height=2.4in,width=3.5in]{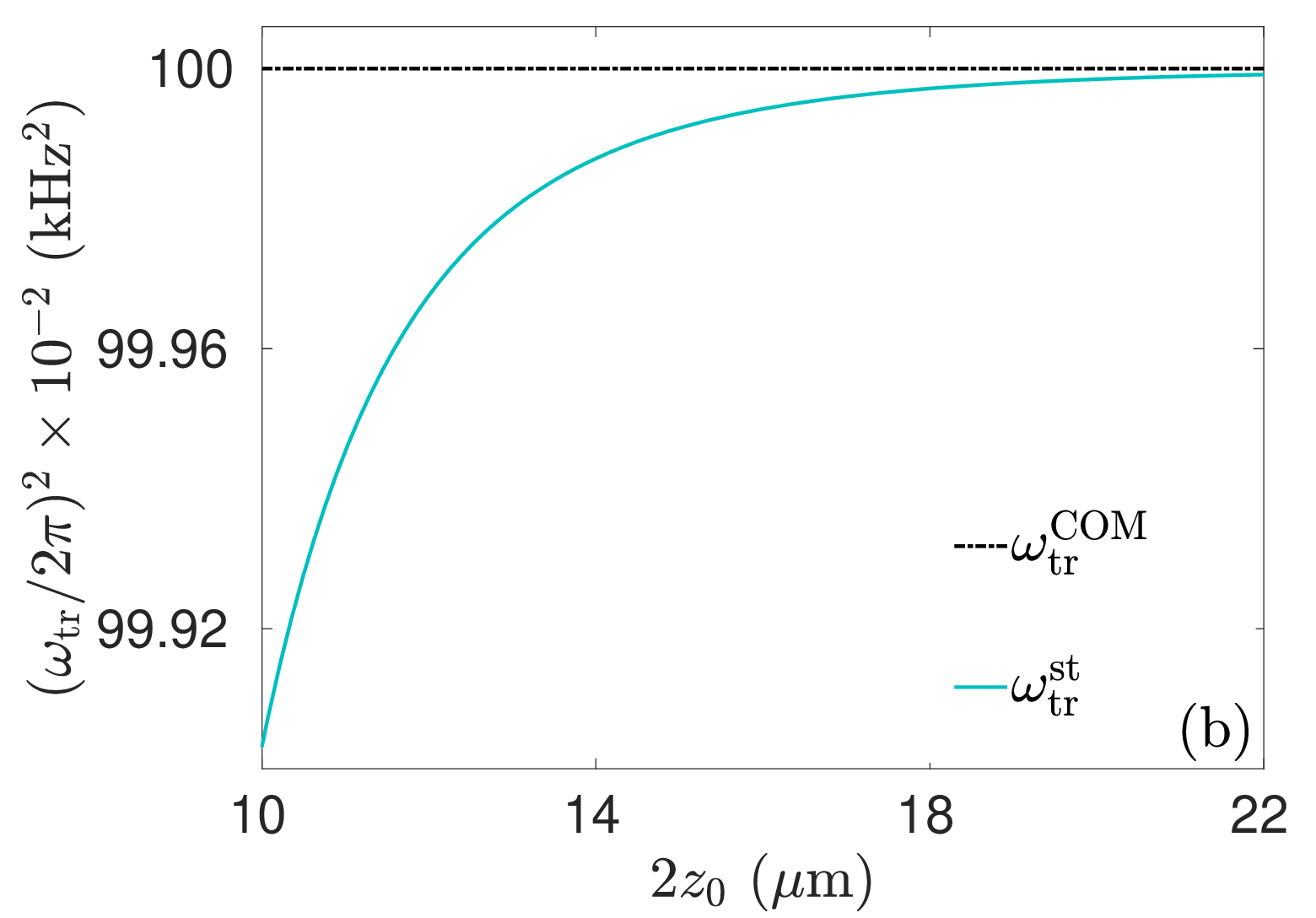}
  \caption{Same as in Fig. \ref{fig 4} but when one atom is in the Rydberg $30S$ state and other is in the ground $5S$ state.}
  \label{fig 5}
\end{figure}

For numerical illustrations, we consider {$^{87}$Rb+$^{40}$Ca$^{+}$+$^{87}$Rb} system where  $^{40}$Ca$^{+}$ ion is trapped at the center of a linear Paul trap and the two $^{87}$Rb atoms reside in two separated, identical optical tweezers. We assume that the tweezers beams are coaxial along the $z$-axis of the Paul trap. Taking the ion-trap's center as the origin of the coordinate frame, the two tweezers's trap centers are located at  $ x_{10} = y_{10} = 0, z_{10} = z_{0}$ and $x_{20} = y_{20} = 0, z_{20} = - z_{0}$. 
We set the radial and axial harmonic trapping frequencies of the ion at  $\omega_{i\rho} = 2\pi \times 1$ MHz and  $\omega_{iz} = 2\pi \times 0.2$ MHz, respectively; which are realistic values \citep{Idziaszek:PRA:2007, Suchorowski:PRA:2022}. The harmonic trapping frequencies of the two optical tweezers in radial and axial directions  are taken to be $\omega_{a\rho} = 2\pi \times 100$ kHz and $\omega_{az} = 2\pi \times 9$ kHz,
respectively. We assume that the atomic qubit is composed of the ground state $|g \rangle = |5 S, F= 1 \rangle$ and the Rydberg state $|r \rangle = |30 S, F = 0 \rangle$, where $F$ represents hyperfine quantum number. These two atomic states may be coupled by a two-photon process via a $P$-excited state. These two long-lived atomic states may be co-trapped in the same optical tweezers  using an optical beam with magic wave-length \cite{Zhang:PRA:2011,Lundblad:PRA:2010}.
The  coefficient  of the long-range interaction between the Ca$^{+}$ ion and the Rb atom in the Rydberg state  is $C_{4} = 3.94 \times 10^{7}C_{4}^{0}$ \citep{Kamenski:JPB:2014}, where $C_{4}^{0} $ is the coefficient when {$^{87}$Rb} is in ground state, $C_{4}^{0}=5.46\times10^{-57}$ J-m$^{4}$ \citep{Tomza:RMP:2019}. The value for Van der Waals coefficient of {$^{87}$Rb} is $-h \times 26.61$ MHz-$\mu$m$^{6}$ \citep{Low:JPB:2012}. The axial length scale of the tweezers trap is  $a_{z} = \sqrt{\frac{\hbar}{m_{a}\omega_{az}}} = 0.11$ $\mu$m. The length scales that characterize atom-ion and atom-atom potentials are $R_{ia}^{*} =\sqrt{\frac{2C_{4}{\mu_{ia}}}{{\hbar}^2}}= 1333$ $\mu$m and $R_{aa}^{*} =\left(\frac{2{C_{6}}\mu_{a}}{{\hbar}^2}\right)^{\frac{1}{4}}= 21.92$ $\mu$m, where $\mu_{ia}= \frac{m_{a}m_{i}}{m_{a}+m_{i}}$ and $\mu_{a}= \frac{m_{a}}{2}$ are ion-atom and atom-atom reduced mass, respectively. The value of $\eta$, which is the ratio between the average speeds of atom and ion, is $0.19$ for our system.

Figure \ref{fig 2} shows the 1D BO potential $V_{\mu}^{\alpha\beta}\left(z\right)$ for $\mu=(0,0,0)$ for three cases, $\left(a\right)$ when both the atoms are in Rydberg state $\left(\alpha\beta=rr\right)$, $\left(b\right)$ when one of the atoms is in ground state and other is in Rydberg state $\left(\alpha\beta=rg\right)$ and $\left(c\right)$ both are in ground state $\left(\alpha\beta=gg\right)$. Here $\mu=(0,0,0)$ implies that the ion is in the ground state of the 3D harmonic potential. We notice that, even at separations larger than $20$ $\mu$m, both $V_\mu^{rr}$ and $V_\mu^{rg}$ are of $10$ kHz order while $|V_\mu^{gg}|$ is quite small ($<1$ Hz). As expected, $|V_\mu^{rr}|>|V_\mu^{rg}|$, and $|V_\mu^{rr}|$ exceeds $|V_\mu^{rg}|$ almost by a factor of 2. The Van der Waals potential between the two Rydberg atoms at $z=10$ $\mu$m is $\sim 26$ Hz for $30S-30S$ Rydberg pair.

In Fig.\ref{fig 3} we show the motional ground state probability density for the atom-pairs along the axial direction for three different trap separations. For large trap separation, the BO potential is small and the atoms are in the ground state of the harmonic trap. As the separation decreases the motional states of the two atoms become coupled so that the ground state becomes admixed with the excited states. Further decrease in separation makes the pair of atoms more strongly coupled and the ground state becomes more distorted due to the coupling with the excited states as shown in Fig.\ref{fig 3}(a).  

We plot the square of the phonon modes as a function of the trap separation $2z_0$ in Fig.\ref{fig 4} and \ref{fig 5}. As observed in Eq.(\ref{eq:17}), the axial and transverse phonon modes in relative DOF are $\sqrt{\omega_{az}'^2-\omega_{zz}^2}$ and $\sqrt{\omega_{a\rho}'^2+\omega_{xy}^2}$, and those in COM DOF are $\sqrt{\omega_{az}'^2+\omega_{zz}^2}$ and $\sqrt{\omega_{a\rho}'^2-\omega_{xy}^2}$, respectively for both atoms in same internal sate ($\alpha=\beta$). The terms $\omega_{xy}$ and $\omega_{zz}$ are 2 to 3 orders of magnitude smaller than both $\omega'_{a\rho}$ and $\omega'_{az}$. As a result, the stretched and COM phonon modes are almost degenerate. For the system $^{87}$Rb$-^{40}$Ca$^+-^{87}$Rb with two atoms in $30S$ Rydberg state, the square of the axial phonon frequency $(\omega_{ax}/2\pi)^2$ is shifted more than 40 kHz$^2$ which is quite significant. However, as the interaction is weaker for the atoms in $25S$ Rydberg state, the shift is relatively small ($<20$ kHz$^2$). Figure \ref{fig 4}(b) shows the variation of the square of the transverse phonon frequency $(\omega_{tr}/2\pi)^2$. The shift in the transverse modes from the asymptotic value of $10^4$ kHz$^2$ ($=\omega_{a\rho}^2/4\pi^2$) is much smaller as compared to the shift in $(\omega_{ax}/2\pi)^2$ owing to the choice of the quasi 1D trap geometry of our system.

For the case where one of the atoms is in ground state and the other in Rydberg state, the stretched and COM phonon frequencies are far from degenerate since ion-Rydberg atom interaction strength is larger than the ion-ground state atom by several orders of magnitude.
As a result the COM mode is almost unchanged in both axial and transverse directions, i.e., $\omega_{ax}^{COM}\approx\omega_{az}$ and $\omega_{tr}^{COM}\approx\omega_{a\rho}$. However, as shown in Fig.\ref{fig 5}, both axial and transverse stretched modes are largely shifted from the asymptotic value as $2z_0$ decreases. It is worth mentioning that there exists a critical value of $2z_0$ below which the system becomes unstable and the corresponding phonon frequencies become imaginary. For the $^{87}$Rb$-^{40}$Ca$^+-^{87}$Rb system, the critical value of trap separation is $2z_0 = 9.20$ $\mu$m for $30S-30S$ pair, $2z_0 = 9.19$ $\mu$m for $30S-5S$ pair and $2z_0 = 7.58$ $\mu$m for $25S-25S$ pair.

From the foregoing analysis of the phonon modes, we have noticed that while the axial phonon modes
depend on the internal electronic states of the two atoms, the transverse phonon are very little sensitive to the internal atomic states. This behavior of the phonon is reflective of the particular geometry that we have chosen for our numerical works. We have assumed a geometric scheme with $z$-axes of the linear ion trap and the two quasi-1D cylindrical atom traps being co-linear. But it is possible to consider other geometry, such as, atom traps’ $z$-axes are not co-linear with but parallel to the $z$-axis of the ion-trap and atom traps are either quasi-2D or 3D. In that case the transverse phonon will also be sensitive to the internal states of the atoms.

Before ending this section, we wish to discuss briefly the consequences of the gauge field or vector
potential that arises due to non-adiabatic effects in the system. This is akin to non-adiabatic effects in molecular physics. In the present scenario, the ion and the atoms play the role of electron and nuclei of a molecule, respectively. This analogy indicates that an ion-atom hybrid system may be useful to explore some aspects of molecular physics. It may be possible to create a conical intersection by coupling two adiabatic BO potentials by driving the ionic qubit with Raman pulses. Then by changing the transverse phonon states of the atoms over a cyclic path by Raman pulses, it may be possible to create geometric phase in the system. Recently, conical intersection has been experimentally studied using trapped ions by optically manipulating spin-phonon coupling \cite{Whitlow:NatChem:2023}. Although, there are observations of conical intersections in molecular physics, the detection of the associated geometric phase in a molecule remains elusive. In this context, our work opens the possibility to simulate the molecular conical intersection and the geometric phase using an ion-atom hybrid system.

\section{Conclusions} \label{Sec:4}
In conclusion, we have proposed a model for creating ion-mediated interaction and phononic coupling
between two distant atoms which are otherwise almost non-interacting. This interaction results from the Rydberg excitation of the atoms. We have used Born-Oppenheimer approximation to separate out the
much faster ionic degrees-of-freedom from the the much slower atomic ones, allowing us to calculate the atom-atom BO potentials. We have analyzed in some detail the characteristics of the axial and transverse phonon modes of the atoms. We have also discussed non-adiabatic effects which lead to a gauge structure and geometric phase in the system. Since this geometric phase arises within the degenerate sub-space of an adiabatic potential, it will entangle the two atoms with the ion. This will result in an intriguing quantum dynamics in the system. Since the phonon states can be optically manipulated for a two-qubit gate operation, our study may be important for quantum computing in such hybrid quantum platform. Furthermore, the ionic qubit may be used as quantum memory to store the outcome of a quantum gate operation. Our system my be scaled up by adding more trapped single atoms arranged in different geometric order around the trapped ion. Such cluster of atoms around an ion may be used as a node for a quantum network or distributed quantum computing using photonic quantum channels to interconnect the nodes. The biggest challenge in realizing such system is to design an electric potential landscape in an ion trap (not necessarily the Paul trap, but other chip-based trap architecture may also be considered) with some positions where the electric field produced by  the ion-trap's electrodes is  small and homogeneous. Then  the optical tweezers can be placed at or near
those positions for the stability of the structure. 

\section{Acknowledgements}
One of us (Subhra Mudli) is grateful to the Department of Science \& Technology, Govt. of India, for  DST INSPIRE fellowship. BD is thankful to Dr. Rajibul Islam, University of Waterloo, and to Dr. Saikat Ghosh, IIT Kanpur, for helpful discussions.


\end{document}